# Criticality and magnetic phases of Ising Shastry – Sutherland candidate holmium tetraboride

Guga Khundzakishvili,[1] Bishnu P. Belbase,[1] Pravin Mahendran,[1] Kevin Zhang,[1,2] Hanjing Xu,[2] Eliana Stoyanoff,[2] Joseph G. Checkelsky,[4] Yaohua Liu,[5] Linda Ye,[3,6] Arnab Banerjee[1]

[1]*Department of Physics and Astronomy, Purdue University, West Lafayette, Indiana 47907, USA*

[2]*Department of Computer Science, Purdue University, West Lafayette, Indiana 47907, USA*

[3]*Department of Physics, Stanford University, Stanford, California 94305, USA*

[4]*Department of Physics, Massachusetts Institute of Technology, Cambridge, Massachusetts 02139, USA*

[5]*Neutron Scattering Division, Oak Ridge National Laboratory, Oak Ridge, Tennessee 37831, USA*

[6]*California Institute of Technology the division of Physics, Mathematics and Astronomy, Pasadena, California, 91125, USA*

**Abstract**: Frustrated magnetic systems arising in geometrically constrained lattices represent rich platforms for exploring unconventional phases of matter, including fractional magnetization plateaus, incommensurate orders, and complex domain dynamics. However, determining the microscopic spin configurations that stabilize such phases is a key challenge, especially when in-plane and out-of-plane spin components coexist and compete. Here, we combine neutron scattering and magnetic susceptibility experiments with simulations to investigate the emergence of field-induced fractional plateaus and the related criticality in a frustrated magnet holmium tetraboride ($HoB_4$) that represents the family of rare earth tetraborides that crystalize in a Shastry-Sutherland lattice in the *ab* plane. We focus on the interplay between classical and quantum criticality near phase boundaries as well as the role of material defects in the stabilization of the ordered phases. We find that simulations using classical annealing can explain certain observed features in the experimental Laue diffraction and the origin of multiple magnetization plateaus. Our results show that defects and out of plane interactions play an important role and can guide the route towards resolving microscopic spin textures in highly frustrated magnets.

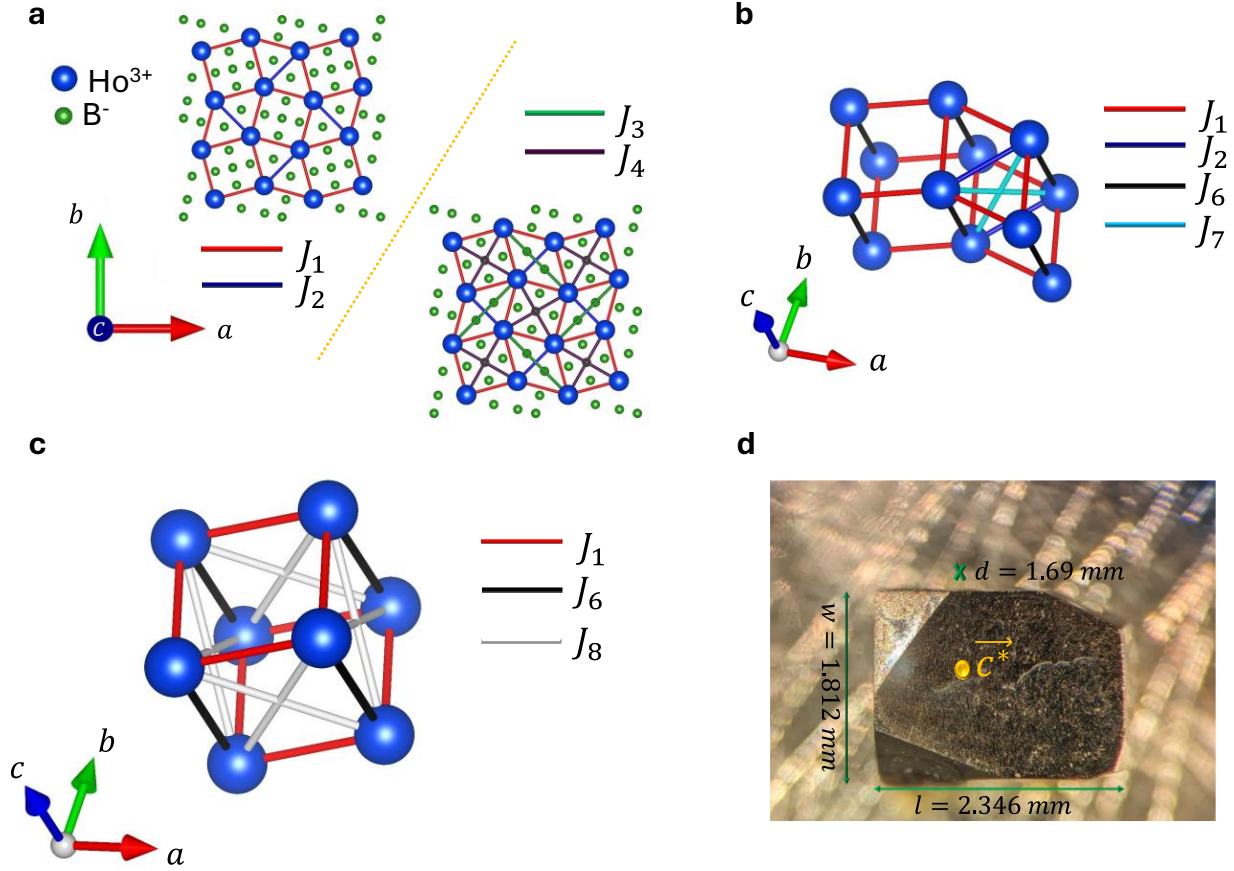

*Figure 1. Lattice of Holmium atoms in $HoB_4$ single crystal. **a**. In-plane structure of $HoB_4$ single crystal. Top corner corresponds to standard SSL lattice with only $J_1$ and $J_2$ interactions. Bottom corner displays additional in-plane $J_3$ and $J_4$ interactions. $J_5$ interaction is not shown here but depicted on Fig. S1. **b**, **c**. Out-of-plane $J_6$, $J_7$, $J_8$ interactions. **d**. single crystal $HoB_4$ with the mass of $m = 61.2$ mg.*

## 1. Introduction

Frustrated magnetic materials exhibit some of the most exotic and poorly understood phases in condensed matter physics, driven by competing interactions that prevent conventional magnetic order [1-3]. Lattices such as the Shastry-Sutherland lattice (SSL), Kagome, and triangular lattices host a variety of unconventional phases, including fractional magnetization plateaus and incommensurate orders [4-6]. Understanding the microscopic spin configurations responsible for these emergent phases, and their evolution with external tuning parameters such as magnetic field and temperature, remains a central challenge in frustrated magnetism [7]. These questions are not only of fundamental interest but are also directly relevant to modern efforts in designing functional quantum material based platforms for emergent phenomena like topological order and quantum spin liquids [8,9].

Among these systems, rare-earth-based SSL magnets HoB₄, TmB₄ and ErB₄ provide an attractive experimental platform, with their strong single-ion anisotropy combined with large magnetic moments [10-12]. These systems exhibit a series of fractional magnetization plateaus, such as $\frac{M}{M_{sat}} = \frac{1}{2}, \frac{1}{3}, \frac{1}{7}, \frac{1}{9}$ under applied field, reflecting non-trivial spin textures stabilized by geometric frustration [13,14]. However, early studies often interpreted these in terms of simple collinear out-of-plane spin configurations [15]. Yet growing experimental evidence, such as the studies of neutron diffraction and magnetometry, suggests a much richer picture involving in-plane spin components, domain formation and complex stripe-like orders [16,17].

Additionally, many frustrated systems exhibit signatures of criticality — both classical critical fluctuations near finite-temperature phase boundaries and quantum criticality near field-induced transitions at low temperatures [18,19]. Disentangling the role of classical vs. quantum critical phenomena in these frustrated lattices and understanding how they connect to the formation of fractional plateaus, and the role of domain dynamics, remains an open question [20,21]. It is important to understand whether a new phase appears from quantum origins or from sub-leading terms in Hamiltonian or perhaps, defects. Such insights are crucial for understanding how magnetically frustrated systems respond to external fields and how they can be tuned toward novel quantum phases.

In this work, we combine field-dependent neutron scattering and simulated annealing to study the emergence of fractional plateaus and critical phases in a single crystal of the frustrated Shastry-Sutherland candidate $HoB_4$. We explore the field-temperature phase diagram, focusing on the interplay between in-plane and out-of-plane order, domain dynamics, and new features that emerge near phase transitions. We discover a new phase – the only one with in-plane magnetic Bragg reflections in $HoB_4$ – close to the critical (C) phase boundary of the ordered antiferromagnetic (AFM) and high temperature paramagnetic (PM) phases which are unreported in literature. From the computationally derived spin patterns, we reveal that in-plane order is intricately affected by the inter-layer interactions, especially close to a phase transition, which are qualitatively similar to the features observed in $HoB_4$ neutron scattering data. Our results show the qualitative effects of the non-ideal terms of the Hamiltonian – the out-of-plane magnetic superexchange interactions and the defects – on the in-plane order in a frustrated spin system. This path presents a unique approach for resolving microscopic order in complex magnetic systems, with implications for a broad class of frustrated quantum magnets. Our results point to complex ordered phases and a delicate phase diagram affected by a variety of terms in the Hamiltonian close to the phase transition which are difficult to explore just using analytical techniques. We propose this system as an excellent candidate for developing Hamiltonian discovery kernels driven by Machine Learning (ML) platforms for Hamiltonian discovery and inverse scattering problems [22, 23].

## II. Shastry-Sutherland model, HoB$_4$ crystal synthesis and structure

The Shastry-Sutherland model is a paradigmatic model in quantum magnetism because of its exact solvability. In this model, the spins are arranged on the square lattice with additional diagonal bonds on alternate plaquettes (red and blue lines for $J_1$ and $J_2$ bonds correspondingly on Fig. 1a), which can be expressed with following Hamiltonian:

$$\widehat{H}_{SSL} = J_1 \sum_{i,j} S_i S_j + J_2 \sum_{k,l} S_k S_l \tag{1}$$

The first summation corresponds to the nearest neighbors, while the second one corresponds to diagonal bonds on alternate plaquettes. This additional arrangement of the alternate plaquettes introduces geometric frustration. This exact solution provides profound insights into the interplay between frustration and quantum fluctuations in low-dimensional quantum systems [24]. The Heisenberg variant of the SSL model encompasses comprehensive quantum spin interactions and accommodates intricate states like dimer-singlet ground states, magnetization plateaus, thermal Hall plateaus and even quantum spin liquids under specific conditions [25-27] and is studied extensively experimentally for the search of exotic plaquette phases and magnetization plateaus especially in SrCu$_2$(BO$_3$)$_2$ [28-31]. The Ising cousin of this model is easier to compute given the commuting Ising spins which makes it a simpler first system to understand intricate details of out-of-plane interactions and defects. Examples of the Ising model include TmB$_4$, NdB$_4$ and HoB$_4$. Conversely, the Ising model - where spins are confined to distinct orientations - provides a computationally manageable approach to comprehending the effects of frustration, defects, and notably out-of-plane interactions, which pose challenges in the complete quantum framework. Compounds such as TmB$_4$ and NdB$_4$ effectively realize this Ising limit, exhibiting magnetization plateaus and commensurate spin structures. In this regard, HoB$_4$ emerges as a candidate system: it exhibits intermediate behavior, merging robust Ising anisotropy with supplementary degrees of freedom from its substantial angular momentum (In the absence of spin-orbit coupling (SOC) and crystal electric field (CEF) splitting J = 8 for Ho$^{3+}$), positioning it as a promising platform for connecting the Ising and Heisenberg interpretations of the SSL and for examining field-tuned classical and quantum criticality.

| T (K) | 296 | | | | | |
|---|---|---|---|---|---|---|
| Crystal system | Tetragonal | | | | | |
| Space group, Hall group | $P_4/mbm, -P\,4_2ab$ | | | | | |
| a(Å), b(Å), c(Å) | 7.01, 7.01, 4.01 | | | | | |
| $\alpha, \beta, \gamma$ | 90°, 90°, 90° | | | | | |
| Volume(Å³) | 202.16 | | | | | |
| Atom | Wyckoff | x | y | z | $U_{iso}$(Å²) | occupancy |
| Ho | 4g | 0.18197 | 0.68197 | 0.00000 | 0.006 | 1.00 |
| B | 8j | 0.32340 | 0.46120 | 0.50000 | 0.007 | 1.00 |
| B | 4e | 0.50000 | 0.50000 | 0.20220 | 0.006 | 1.00 |
| B | 4h | 0.08700 | 0.41300 | 0.50000 | 0.008 | 1.00 |

*Table 1: single crystal X-ray data of HoB₄*

Single crystals of HoB$_4$ have been synthesized using floating zone method. Polycrystalline Ho$^{11}$B$_4$ was first synthesized by reacting thoroughly mixed and compressed Ho$_2$O$_3$ and $^{11}$B$_4$ powder in an argon gas flow. The resulting polycrystalline rods were then zone-refined to obtain high quality single crystals. To determine the crystal structure, single-crystal XRD data were collected in single crystal X-ray laboratory in Chemistry department of Purdue University using a Bruker D8 Quest diffractometer equipped with Mo Kα radiation ($\lambda = 0.71073$ Å). Measurements were performed at room temperature. The results of single crystal XRD is presented in table 1. The crystal magnetic lattice in 2D is shown in Fig. 1a-c and conforms to an SSL with lattice constants $a = b = 7.01$ Å and $c = 4.01$ Å. The consequent reciprocal lattice vectors are $a^*, b^*, c^*$ are $a^* = b^* = \frac{2\pi}{a} = 0.88$ Å$^{-1}$ and $c^* = \frac{2\pi}{c} = 1.57$ Å$^{-1}$ respectively.

In Fig. 1a-c the SSL interactions are denoted by J$_1$ and J$_2$ providing the primary mechanism for frustration which makes the model topologically equivalent to the Shastry Sutherland model. Additional interactions, which deviate from the simple Shastry-Sutherland type, are also expected to be present in a realistic situation, which are also represented in Fig. 1a-c such as additional in-plane interactions $J_3 - J_5$ and out-of-plane $J_6 - J_8$ which are discussed more in section V.

A small number of defects and stacking faults can change the exact nature of the phases especially close to criticality as well as the phase diagram in 2D crystals. The exact nature of the phase diagram is often found to be sample dependent, and it is difficult to reconcile measurements performed using different techniques on different samples – especially for the delicate phases. Additionally, for large samples with a large effective magnetic moment, the demagnetization factors are different for different sample shapes making it hard to reconcile measurement and phase diagrams from different measurements. Unlike previous measurements on $HoB_4$, for this manuscript, we have performed all measurements, both neutrons diffraction and susceptibility/magnetization measurements on the same single crystal of $HoB_4$ (Fig. 1d) allowing us a direct comparison of the results.

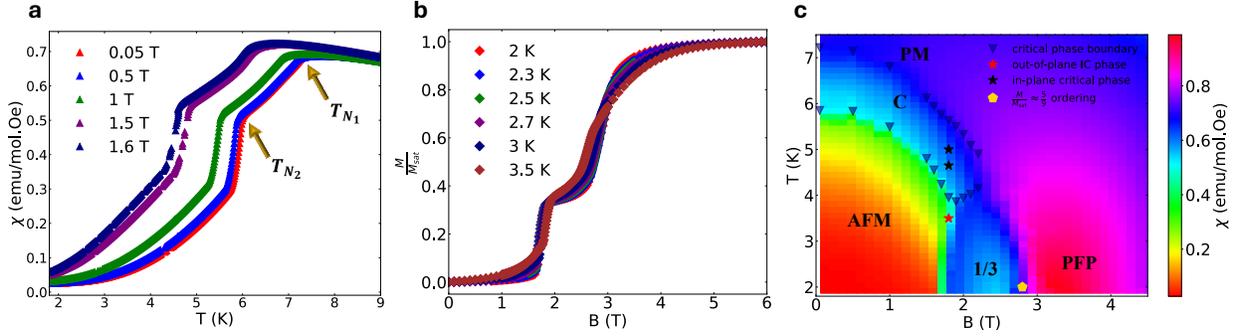

*Figure 2: **a**. Magnetic susceptibility as a function of temperature displaying two consecutive phase transitions between paramagnetic (PM) and low temperature antiferromagnetic (AFM) phases. Fig. S2 displays detailed phase transition analysis for applied $B = 0.05\ T$ magnetic field. **b**. Magnetization normalized with saturation magnetization as a function of applied field, displaying $1/3$ magnetization between AFM on low fields and partially field polarized (PFP) phases on high fields. 2 K magnetization data on Fig. S3 depicts the trace of additional magnetization ($M/M_{sat} \approx 5/9$) that appears between $1/3$ and PFP phases. **c**. Magnetic phase diagram of $HoB_4$ reconstructed from magnetic susceptibility measurement, displaying following phases: AFM, PM, PFP, $M/M_{sat} = 1/3$, $M/M_{sat} \approx 5/9$ and in-plane classical critical phase (C) that supports the in-plane order (details in text).*

### III. Magnetic Susceptibility measurements

Magnetization/Susceptibility measurements are a staple technique to divulge the magnetic phases and magnetic phase transitions. To deduce the phase diagram and critical regimes in $HoB_4$, we performed magnetic susceptibility measurements using the Superconducting Quantum Interference Device (SQUID) magnetometer (MPMS-3 Quantum Design), equipped with [4]He insert at the BIRCK Nanotechnology Center in Purdue University with a base temperature of $T = 1.8$ K. The measurement has been conducted in two separate parts: **1.** Magnetic field sweep for fixed temperature and applied magnetic field along $c$ ($c^*$) axis, yielding $M$ vs. B dependence; Magnetization value is normalized as $M \to M/M_{sat}$, where $M_{sat}$ represents saturated magnetization (in the units of $\frac{\mu_B}{Ho^{3+}}$) for applied saturation magnetic field. **2.** Temperature sweep for the fixed applied magnetic field along $c$ ($c^*$) axis, extracting magnetic susceptibility $\chi = \frac{M}{H}$ vs. $T$.

The sensitive nature of phase diagram in $HoB_4$ is immediately apparent in magnetic susceptibility which provides us with additional information regarding phase transitions. The analysis of phase transitions is split into several different regions. As shown on Fig. 2a, by the evolution of magnetic susceptibility at zero field, the sample undergoes two discontinuities at $T_{N_1} = 7.22$ K and $T_{N_2} = 5.97$ K. The $T_{N_1}$ and $T_{N_2}$ vary with the application of the out-of-plane magnetic field, clearly defining the boundary of the low temperature phase. The low temperature phase shows signatures of an antiferromagnetic (AFM) spin arrangement, as inferred from the decreased susceptibility of the sample to the applied magnetic field. The phase transition at $T_{N_1}$ (between high temperature paramagnetic and intermediate critical phase) displays broader maximum pointing towards 2$^{nd}$ order in nature. Conversely, for the phase transition at $T_{N_2}$, magnetic susceptibility experiences a sharp drop which suggests a likely 1$^{st}$ order phase transition to AFM phase.

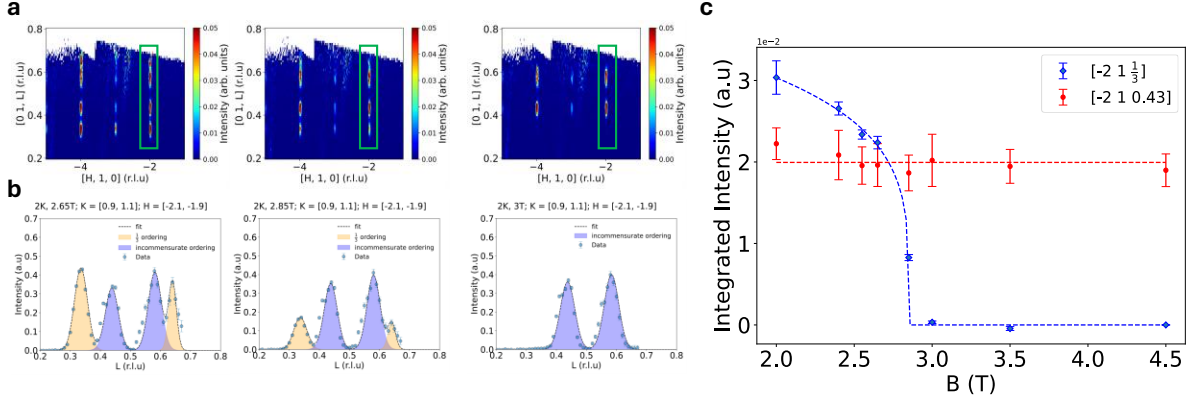

*Figure 3: Neutron diffraction data in (H 1 L) planes. **a**. evolution of 1/3 and incommensurate orders in the range of applied field 2.65 – 3 T; 2D slices have been made for $K = [0.9, 1.1]$ r.l.u integration range. **b**. Evolution of 1/3 and incommensurate orderings in the range of applied field 2.65 T – 3 T; 1D slices have been made for $K = [0.9, 1.1]\ r.l.u$ and $H = [0.9, 1.1]\ r.l.u$ integration range. **c**. Integrated peak intensities as the function of applied magnetic field.*

The results are summarized in Fig. 2 on phase diagram constructed from the combination of all the magnetization measurements. It reveals the paradigmatic features of a quantum phase transition with the following phases: (1) AFM phase at low fields (2) a 'partially field polarized' (PFP) phase at high fields. (3) The 1/3$^{rd}$ magnetization phase emerging close to the critical point of the AFM→PFP phase transition. The triangles, which represent phase transition temperatures and fields derived from susceptibility analysis, outline the boundary of classical critical (C) phase sandwiched between $T_{N_1}$ and $T_{N_2}$ that appears above the phase boundary of the AFM phase. The 1/3$^{rd}$ magnetization plateau forms at around B = 1.7 T and sustains up until B = 2.8 T (Fig. 2b). As we would discuss in the next section, we find this phase coexists with an incommensurate (IC) spin ordered phase which sustains to even higher temperatures and fields, and the phase marked PFP supports only the incommensurate (IC) order which exists everywhere except for in the AFM phase.

The phase diagram observed by us in the susceptibility measurements is simpler than the one proposed earlier. Besides the 1/3$^{rd}$ plateau in the vicinity of the AFM→PFP transition, an earlier source [32] had claimed that additional magnetization plateaus can be observed in the narrow ranges of magnetic field for the temperatures at and below 2 K, such as $\frac{M}{M_{sat}} = \frac{1}{2}; \frac{4}{9}; \frac{3}{5}$ magnetization plateaus. Our magnetization measurements conducted on single crystal HoB$_4$ instead just reveals a trace of a plateau at roughly $\frac{M}{M_{sat}} = 0.55 \cong \frac{5}{9}$ between 1/3$^{rd}$ plateau and PFP phase (Fig. S3), which is marked with golden pentagon on Fig. 2c. Overall, comparing our data with [32], we can also conclude that some of the plateaus could be non-universal, and sample dependent. Whether these are stabilized in defects and different out-of-plane interactions, or quantum fluctuations are possibilities we will address more in section V.

## IV. Neutron diffraction

Neutron Laue diffraction provides the static structure factor which is a robust tool for determining phases of magnetic materials, revealing magnetization plateaus and commensurate/incommensurate magnetic orders. It is a robust tool for understanding the delicate phases that arise at critical regimes of a phase diagram. Neutron diffraction measurements on a single crystal HoB$_4$ (Fig. 1d) were performed at Spallation Neutron Source (SNS), Oak Ridge National Laboratory (ORNL) at the Elastic Diffuse Scattering Spectrometer CORELLI with the incident neutron energy range of $E_i = 10 - 200$ meV. CORELLI employs a white-beam Laue diffraction method, allowing for simultaneous collection of diffraction data across a broad range of wavelengths. To discriminate between elastic and inelastic scattering, CORELLI utilizes a statistical chopper that modulates the incoming neutron beam quasi-randomly. The resulting data are processed using a cross-correlation method, which reconstructs the elastic scattering signal by correlating the modulated incident beam with the detected scattered neutrons. This approach effectively isolates elastic components from the total scattering data. The sample was aligned in the $ab$ plane (i.e., in the 2D Shastry-Sutherland) plane with the magnetic field pointing parallel to the $c$ ($c^*$) axis and rotated in the range of $\pm 60°$ around $c$ axis, with the step size of $2°$.

We first concentrate on the location of the 1/3$^{rd}$ magnetization plateau in the field regime $B = 2.65 - 3.0$ T at $T = 2$ K. Fig. 3a shows the data on the $[H\ K = 1\ L]$ slice, where $K$ is integrated between [0.9, 1.1] r.l.u. We immediately notice that the magnetic Bragg peaks remain at integer locations along $[H, K]$ i.e., on the $ab$ plane. However, in the out-of-plane, or the stacking direction we reveal peaks at fractional indices: the ferrimagnetic phase $L = \frac{1}{3}$ r.l.u and incommensurate phase with $L_{inc} = (0.43 \pm 0.012)$ r.l.u on Fig. 3a. The results are clarified further in Fig. 3b where we provide a cut along $[-2\ 1\ L]$ that is fitted to two Q-vectors at $L = 0.43$ r.l.u and $L = 0.33$ r.l.u (and their Fourier conjugate peaks at $L = 0.57$ r.l.u and $L = 0.67$ r.l.u). We observe no additional magnetic Bragg peaks in c direction in neutron scattering. Most of the features are consistent with the 1/3$^{rd}$ and the 0.43 orders that were reported earlier [32]. We performed a meticulous search for the additional phases such as narrow magnetization plateaus at $\frac{M}{M_{sat}} = \frac{1}{2}; \frac{4}{9}; \frac{3}{5}$ reported earlier and found the trace of none. However, the trace of $\frac{M}{M_{sat}} \approx \frac{5}{9}$ ordering has been established at $T = 2$ K (Fig. S3). Thus, it is reasonable to conjecture that the existence of certain magnetization plateaus in HoB$_4$ can be specific to the sample and could arise in the presence of non-universal defects, as is discussed in section V of this study.

Figure 3c depicts the evolution of intensities of magnetic Bragg peaks related to out-of-plane magnetic orders. The scattering intensities for incommensurate Bragg peak [-2 1 0.43] sustains to large magnetic fields, while ferrimagnetic commensurate ordering [-2 1 1/3] diminishes as the function of applied magnetic field and fully disappears in the vicinity of $B = 3$ T.

Notably, the integrated intensity of the incommensurate phase is unaffected through the diminishing of the 1/3$^{rd}$ order leading into the partially field polarized phase at 3 T (Fig. 2c). This is particularly surprising given that HoB$_4$ has only one Holmium site, with the overall ordered moment conserved and hence it is natural to expect a competition between long-range orders and a trade-off between net ordered moments of different Q-vectors. Our results instead show that the IC phase is robust and derives from mechanisms (or terms in the Hamiltonian) distinct from the 1/3$^{rd}$, C, and PFP phases.

In the next section, we concentrate on the regime of classical criticality intermediate between $T_{N_1}$ and $T_{N_2}$ in the susceptibility phase diagram in Fig. 2c. Our motivation was driven by a goal to discover new phases close to the critical regime, where enhanced effects of long-range fluctuations can lead to the genesis of new in-plane order where sub-leading terms of the Hamiltonian play a more defining role. In-plane indexed peaks have been observed in other rare-earth SSL candidates such as NdB$_4$ [14] and TmB$_4$ [15].

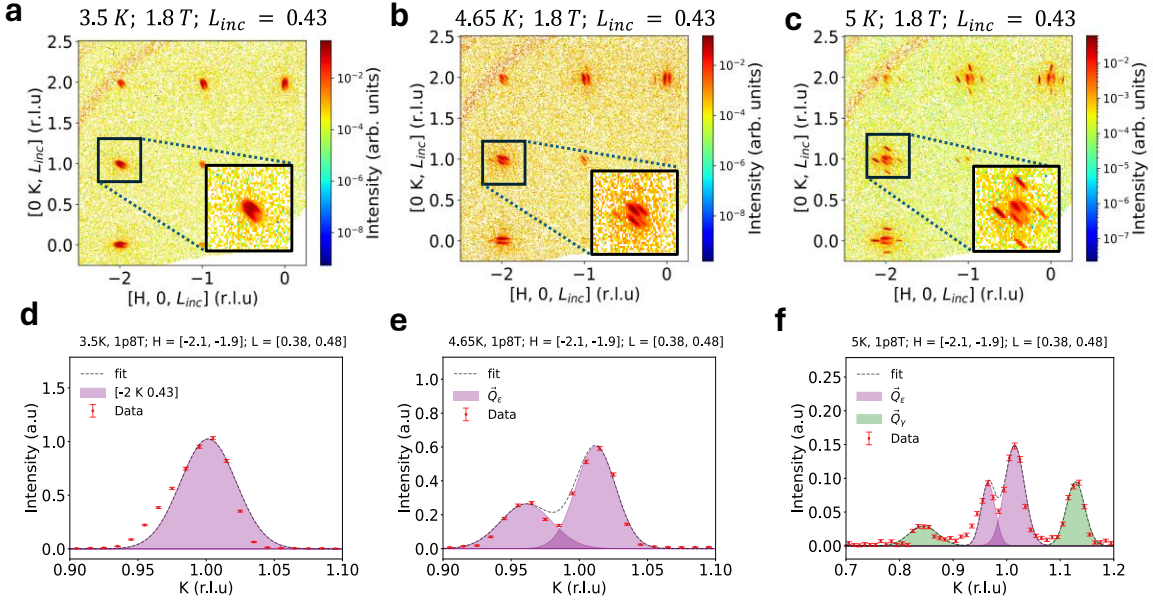

*Figure 4: In-plane ordering of critical intermediate phase. 2D slice of $(H\ K\ L_{inc})$ has been made with $L_{inc} = (0.43 \pm 0.05)$ r.l.u integration range at 1.8 T field and [3.5, 4.65, 5] K temperatures. **a-c**. emergence of $(0, \pm\varepsilon, \delta')$ and $(0, \pm\gamma, \delta')$ ordering of incommensurate peak splitting as the function of temperature. **d-f**. 1D cut on K reciprocal direction for $H = (-2 \pm 0.1)$ r.l.u integration range. $(0, \pm\varepsilon, \delta')$ ordering with $\varepsilon = (0.0245 \pm 0.004)$ r.l.u $\delta' = 0.43$ r.l.u and $(0, \pm\gamma, \delta')$ with $\gamma = 0.14$ r.l.u. $\frac{1}{7}$ ordering exists in the vicinity of 5 K, 1.8 T and is absent for 4.65 K, 1.8 T. Incommensurate out-of-plane ordering sustains throughout the entirety of intermediate phase between $T_{N_1}$ and $T_{N_2}$.*

The critical phase turns out to be good place to find for such phases in HoB$_4$ because strong fluctuations could enhance the delicate sub-leading terms of the Hamiltonian.

At the critical regime (black stars in Fig. 2c), we discover a new phase corresponding to magnetic Bragg peak splitting into in-plane commensurate and incommensurate orderings. The transition into the critical phase is further explained in the slices of $(H\ K\ 0) \perp c^*$ reciprocal planes for $B = 1.8$ T. As it has been shown in in-plane neutron diffraction measurements on Fig. 4a-c, upon increasing the temperature from 3.5 K to 5 K, we report two separate in-plane orders: (1) the evolution in-plane magnetic order that results to $[H\ K\ L_{inc}]$ splitting into $\vec{Q}_\varepsilon = (\pm\varepsilon, 0, \delta')$ and $\vec{Q}_\varepsilon = (0, \pm\varepsilon, \delta')$ satellite peaks along $H$ and $K$ directions correspondingly (Fig. 4b), where $\varepsilon = (0.0245 \pm 0.004)$ r.l.u and $\delta' = 0.43$ r.l.u. This is accompanied by another new set of satellite peaks with the modulation vectors $\vec{Q}_\gamma = (0, \pm\gamma, \delta')$ and $\vec{Q}_\gamma = (\pm\gamma, 0, \delta')$ at $T = 5$ K (Fig. 4c). We fit the cut along $K$ at the $[-2\ K\ 0.43]$ in Fig. 4f leading to $\gamma = 0.142 \pm 0.00025 \approx \frac{1}{7}$ r.l.u. Thus, at the critical point we report on the discovery of a, likely commensurate, in-plane $H = K = 1/7$ ordering. The corresponding modulation vector for the order is: $\vec{Q} = \vec{Q}_\varepsilon \oplus \vec{Q}_\gamma$. We note that this is the only in-plane order in HoB$_4$, appearing as an additional modulation of the incommensurate order along $c^*$.

The signature of field-induced $(\pm\gamma, \pm\gamma, \delta')$ ordering with in-plane indices exists in the very narrow range or temperature and exists in the critical phase shown in section III. If such a splitting of in-plane ordering can be triggered by out-of-plane interactions between $Ho^{3+}$ ions, which consists of superexchange and dipole-dipole interactions, is qualitatively discussed in the next section.

Overall, the results show that the $L_{inc} = \delta' = 0.43$ r.l.u phase is absent only in the AFM phase, but otherwise particularly robust in HoB$_4$. Robustness suggests that it arises from leading terms of the effective Hamiltonian. The other orders, such as the 1/3$^{rd}$ plateau, are sub-leading effects. The AFM phase trade intensity into the incommensurate order. It is therefore important to deduce nature of the incommensurate phase, using a 3D model where frustration is not only from in-plane interactions, but also arise in strong out-of-plane interactions, before analysis of the other sub-leading orders codependent on the incommensurate phase would be possible. While the exact origin of the $L_{inc} = 0.43$ incommensurate order is still a mystery, a qualitative attempt to understand the origins of several other features of the critical phase is presented next.

## V. SSL simulation

The 2D Ising SSL model is shown in Eq. (1). This 2D SSL model has been analyzed in detail first by Dublenych [33] with no additional interactions, and then in [34] in the presence of additional interactions where new plateaus $\frac{M}{M_{sat}} = \frac{1}{9}, \frac{1}{6}, \frac{2}{3}, \frac{4}{9}, \frac{1}{2}$ with in-plane order were theoretically observed. Additionally, this model represents an appealing platform for studying long-range ordering and continuous and discontinuous quantum phase transitions in 2D frustrated magnets [35-37]. The results are also analyzed and reproduced using D-Wave quantum annealer [38], where the classical ground state and the phase transitions were reproduced, and spin behavior at the critical points predicted. Later in [39] the expanded version of SSL Hamiltonian (1$^{st}$ and 3$^{rd}$ summation of Eq. (2)) was analyzed using the Fujitsu parallel tempering machine to discover a host of new phases, such as the in-plane 5/9$^{th}$ magnetization plateau, that does not exist in the just a $J_1 - J_2$ model. We note that the prior analysis is limited to the 2D Ising SSL which is insufficient for HoB$_4$ since much of the ordered phases observed by us are out-of-plane. The in-plane order which we observed is in the envelope of the out-of-plane incommensurate phase $L_{inc} = 0.43$ r.l.u. also underscoring the importance of a full 3D treatment.

Questions can arise whether: **a)** The features observed in HoB$_4$ can be described by an Ising Hamiltonian, and **b)** whether an SSL Hamiltonian with out-of-plane interactions may be able to capture some of the observed features? To answer the first question, we refer to figures 7 and 9 in ref. [32] and the discussion therein. The AFM phase has a spin canting 23° away from the $c$ axis, thus our field along $c$ axis has a sizeable transverse component making perhaps, a transverse field Ising model more appropriate. On the other hand, the phases which support the IC, C, 1/3$^{rd}$ and the PFP phases all have spin moments along the c axis. Given that our field B || $c^*$ is hence a longitudinal field, these phases observed in HoB$_4$ could be compliant to an Ising SSL model with a longitudinal field.

To answer the second question, while a full treatment of the incommensurate out-of-plane order will require a very large 3D lattice which could be very useful future work, we show that a longitudinal-field Ising 3D SSL model (i.e., a 2D SSL model extended with out-of-plane interaction terms) inspired by the XRD data of HoB$_4$ single crystal, can qualitatively capture some of the features we observed in the neutron scattering data in the critical phase.

We consider an extended SSL model, starting with the results of [39] but with additional out-of-plane interaction terms to seek the behavior of the spin-spin correlations close to a phase transition:

$$\hat{H} = \sum_{i=1}^{5} J_i \sum_{k,l} S_k S_l + \sum_{i=6}^{8} J_i \sum_{k,l} S_k S_l - h \sum_m S_m \qquad (2)$$

The first term including $J_1 - J_5$ represents in-plane interactions depicted in Fig. 1a; The second term including $J_6 - J_8$ is out-of-plane interactions shown in Fig. 1b-c; The out-of-plane interaction $J_6$ is parallel to $c$ axis and represents direct out-of-plane connection between $ab$ planes. $J_7$ and $J_8$ interactions are located under $J_2$ and $J_1$ respectively. We additionally implement a Zeeman coupling of out-of-plane magnetic field with individual spin sites denoted by the last term of Eq. (2) ($m$ runs over the entire spin array). The corresponding magnetic phases can be identified by mapping $M = \langle S_m \rangle$ in $(J_i, h)$ 9-dimensional parameter space, which afterwards gets translated into $M$ vs. $h$ for appropriate $J_i$ values.

| | | | |
|---|---|---|---|
| $r_2/r_1$ | **0.99** | $J_2/J_1$ | **1.03** |
| $r_3/r_1$ | 1.73 | $J_3/J_1$ | 0.19 |
| $r_4/r_1$ | 1.41 | $J_4/J_1$ | 0.35 |
| $r_5/r_1$ | 1.93 | $J_5/J_1$ | 0.14 |
| $r_6/r_1$ | 1.09 | $J_6/J_1$ | 0.77 |
| $r_7/r_1$ | 1.47 | $J_7/J_1$ | 0.314 |
| $r_8/r_1$ | 1.48 | $J_8/J_1$ | 0.308 |

*Table 2: Initial guess of interaction parameters according to dipole-dipole interactions. $r_i$ represents the distances between Holmium ions connected through $J_i$ interaction in $HoB_4$ single crystal as measured using XRD in Section II.*

Starting with spin-spin interaction values $(J_i)$ inspired from experimental data, we perform simulated annealing and parallel tempering on the Neal simulated annealer offered by D-Wave which allows us to explore an Ising Hamiltonian with arbitrary connectivity. We compute ground states of Eq. (2) and compare their magnetization values with the applied transverse magnetic field following the same steps in [33, 34]. The existence of magnetization plateau(s) can be confirmed by plotting magnetization as the function of applied magnetic field for fixed $J_i$ values. The direction of the ordering can be determined by fast Fourier transform (FFT) of extracted spin structure from lattice spin state. Fourier analysis of spin structure yields information about the symmetry of the system of spin arrangements, which qualitatively carries the same information as magnetic Bragg peaks in neutron diffraction measurements. Once the magnetization plateau is determined, we extract the in-plane and out-of-plane spin structure factors via FFT of the two-spin correlation function.

Due to the large magnetic moment of $Ho^{3+}$, dipole-dipole interaction is an important part of overall interactions between spin sites. We extract the initial parameters of magnetic Hamiltonian in Eq. (2) from the interionic distances of $Ho^{3+}$ (table 2) using the inverse cubic nature of magnetic dipole-dipole interaction:

$$U_{d-d} \sim S_1 S_2 / r^3 \qquad (3)$$

The initial target values for spin-spin interactions (which itself includes both dipole-dipole and exchange interactions) thus inferred are presented in table 2 from XRD measurements.

We implement 5 in-plane and 3 out-of-plane spin-spin interactions (Fig. 1a-c) using a lattice size of $60 \times 60 \times 9 = 5400$ spins. Spin-spin interaction parameters are initialized based on table 2 and varied within $\pm 60\%$ range. For visualization purposes we plot separate phase diagrams as colormaps on $J_i$ vs. $h$, for $i \epsilon [2,8]$ ($J_1 = 1$). Analysis of phase diagram(s) reveals magnetization plateau at $|M| = 0.5$ (Fig. 5a) marked with yellow color. In Fig. 5b it is evident that ½ magnetization plateau becomes less pronounced as we increase out-of-plane interaction parameter $J_7$. Corresponding spin arrangement *down-down-down-up* (Fig. 5c) evolves according to Fig. 5d-f represented with FFT calculation of spin ordering in $xy$ plane:

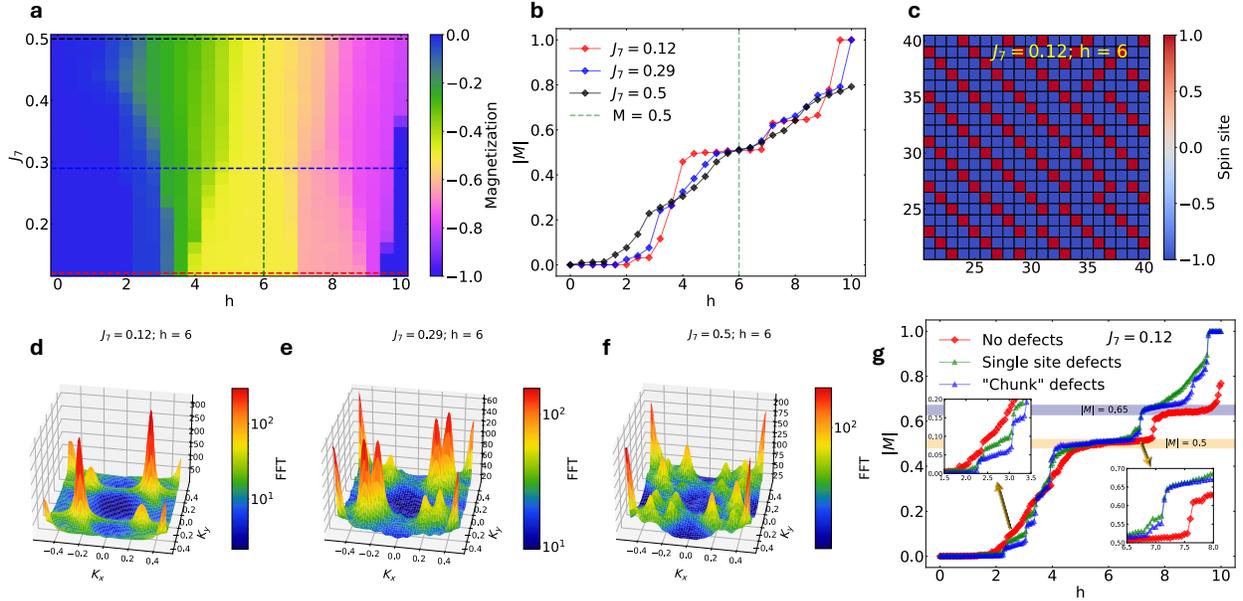

*Figure 5:* ***a***. *Phase diagram for variable $J_7$ and h in the absence of defects. The rest of parameter values: $J_1 = 1$, $J_2 = 1.03$, $J_3 = 0.19$, $J_4 = 0.35$, $J_5 = 0.14$, $J_6 = 0.77$, $J_8 = 0.308$.* ***b***. *The evolution of magnetization plateau $|M| = 0.5$, as the function of $J_7$ in the absence of defects.* ***c***. *in-plane $|M| = 0.5$ magnetic ordering for $J_1 = 1$, $J_2 = 1.03$, $J_3 = 0.19$, $J_4 = 0.35$, $J_5 = 0.14$, $J_6 = 0.77$, $J_7 = 0.12$, $J_8 = 0.308$, $h = 6$, showing spin down-down-down-up ordering for the middle part of layer 5 (the system consists of $60 \times 60 \times 9$ spins).* ***d-f***. *Fourier transform of spin structures for $J_7 = [0.12, 0.29, 0.5]$ displaying the evolution of in-plane magnetic ordering as the function of out-of-plane interaction $J_7$ in the absence of defects.* ***g***. *Magnetization evolution for $J_7 = 0.12$ with and without defects, displaying the appearance of additional magnetization plateaus when single and 'chunk' defects are introduced.*

$$FFT = \tilde{C}(k_x, k_y) = \sum_{x=0}^{N_x-1} \sum_{y=0}^{N_y-1} C(x,y) e^{-i\left(\frac{2\pi x k_x}{N_x} + \frac{2\pi y k_y}{N_y}\right)} \quad (4)$$

Where $C(x,y) = \langle S(0,0), S(x,y) \rangle$ is two-spin correlation function. Since the observed in-plane spin structure consists of *down-down-down-up* ordering (for $J_7 = 0.12$), there are repeating patterns over 4 spins in $xy$ plane, resulting into sharp $k_x, k_y = \pm 0.25$ propagation vectors in $\tilde{C}(k_x, k_y)$ (Fig. 5d).

As out-of-plane interactions increase, these peaks phase into two and consecutively four peaklets. Even though $|M| = 0.5$ plateau is not observed in single crystal HoB$_4$, and our data does not exactly solve its Hamiltonian, the qualitative analysis of in-plane ordering confirms that in-plane features observed in frustrated Shastry-Sutherland magnets can indeed stem out of out-of-plane interaction terms of extended SSL Hamiltonian.

We notice that the treatment can produce both in-plane and out-of-plane orders but fails to reproduce the $L_{inc} = 0.43$ or the 1/3$^{rd}$ phase. The exact replication of in-plane and out-of-plane phases requires thorough search of $J_i$ parameter space. Considering that our evaluations of the values of $J_i$ parameters were based solely on dipole-dipole interaction, the actual interaction values for HoB$_4$ single crystal can be different enough to cause the emergence of different orderings. Yet, we can attract attention to the in-plane splitting of peaks due to increasing out-of-plane couplings, shown in Figs. 5 d-f. These results provide a compelling idea of a scenario that in-plane orderings in SSL magnet can be triggered via out-of-plane interactions.

We further explore the appearances of magnetization plateaus as the result of spin lattice defects, such as missing spin site, or slight $\delta J_i$ shift between certain lattice points. By removing certain spins off the lattice (equivalent to creating $S = 0$ defects arising from vacancies), we study magnetization as a function of magnetic field and check the appearance of additional magnetization plateaus, which are not observed in the absence of defects. Furthermore, we consider two different sizes of defects by removing **(1)** individual spins from arbitrary $(x, y, z)$ locations (we remove overall 1000 spins), we call these single site defects and **(2)** $3 \times 3 \times 3$ size domains from arbitrary $(x \pm 1, y \pm 1, z \pm 1)$ locations (we remove overall 30 domains), we call it 'chunk' defects.

Magnetization evolution as a function of applied field for $J_7 = 0.12$ depicts the appearance of several new magnetization plateaus in the narrow range of magnetic fields (Fig. 5g). For multiple field values such as $h = 2.5$ and $h = 7$, it is apparent from Fig. 5g (inset) that new magnetization plateaus appear when the defects are introduced. These new plateaus appear at the critical regime between $|M| = 0$ and $0.5$ at the locations $|M| \approx 0.05, 0.15$ and $0.25$. Furthermore, introducing the defects also shifts the range of the field values over which a plateau is stabilized, seen most prominently for the $|M| = 0.5$ and $0.65$ magnetization plateaus, which now appear at smaller fields. Defects break frustration. Besides defects having an influence on the formation of the plateaus, larger defects can start to wash out existing plateaus towards a smoother spin-glass like rise as shown for $|M| \approx 0.65$ magnetization plateau with the so-called "chunk" defects. This proves that defects and their sizes play a major role, especially when comparing data from different samples and in deduction of the truly universal features of the data.

Overall, we have shown that the simulated annealing approach of 3D SSL lattice with defects yields qualitative explanations for the origin of some of the features of the in-plane ordering in frustrated SSL magnets. In-plane structure analysis shows that we can manipulate in-plane magnetic ordering by changing out-of-plane interactions, splitting sharp peaks into pairs and quadruplets of smaller peaklets, much as in neutron scattering data analysis in Fig 4. Additionally, we have shown that small magnetization plateaus can appear in the presence of defects, hence can vary from sample to sample.

## VI. Conclusion

This study investigated the magnetic phases and transitions in HoB$_4$ using neutron diffraction, susceptibility measurements, and computational modeling. A critical regime confirmed by susceptibility measurements was found to harbor a new phase with in-plane Bragg peaks in neutron diffraction. A careful analysis of this phase pointed to an in-plane peak splitting with corresponding propagation vector $\vec{Q}_\gamma$ corresponding to a (likely commensurate) $1/7$ ordering. This result showcases the potential to discover new magnetic phases in the critical regions of phase diagram of SSL inspired quantum magnets and perhaps, more generally in geometrically frustrated magnets.

The origins of in-plane ordering have been partially explained with the means of implementing the system as 3D SSL lattice and recovering the phases with the means of annealing algorithms. As it has been shown in Fig. 2c $\frac{M}{M_{sat}} \approx \frac{5}{9}$ appears at $T = 2$ K on magnetic phase diagram. It is interesting to note that a phase with a 5/9$^{th}$ magnetization was also extracted in [39] in the extended 2D Ising SSL model using Fujitsu parallel tempering machine. It remains to be seen in future work whether these phases could be related.

Our analysis of simulated annealing using Neal simulated annealer (offered by D-Wave) indicates multiple closely spaced wave vectors arising from out-of-plane interactions that could be explained within the Ising SSL model extended to 3D. The results look qualitatively similar to the peak splitting we observe in $HoB_4$, although more work is required to find either the $1/7^{th}$ in-plane split peaks or the $1/3^{rd}$ out-of-plane peaks observed in our data.

Furthermore, via the analysis of a defective lattice we suggest a mechanism for the formation of new plateaus and features which has been reported by certain groups earlier close to criticality. It has been established that in-plane magnetic ordering can be affected by out-of-plane interactions and most importantly with the existence of defects which break the frustration. Variation of out-of-plane interactions splits in-plane ordering, producing (qualitatively) similar features as depicted on Fig 4. Introducing the defects in the system shows the appearance of narrow plateaus, otherwise non-existent in the SSL system. This can perhaps explain why in certain samples (possibly with some defects) additional magnetization plateaus can be observed. We have shown that defects can introduce new magnetization plateaus especially at the critical points. In our sample we did not notice many of the small plateaus observed in some previous reports above the $1/3^{rd}$ plateau, which might therefore suggest a sample with lesser defects. This leads to a need for caution when analyzing data from ferrimagnetic samples. Yet, the smaller magnetization plateaus could also provide a unique fingerprint for the internal structure of the crystal which could be extracted using further sophisticated tools and AI-driven techniques.

A quantum phase transition observed in $HoB_4$ with interactions beyond just the longitudinal-field Ising model requires a complete and quantum treatment of the phase transition which requires DMRG techniques which are beyond the purview of the manuscript. A transverse field Ising model (TFIM) can be emulated in a Rydberg atom-based platform such as QuEra [40] or using the fast-quench techniques using D-Wave quantum annealers [41] or using the Google Sycamore platform [42]. Unfortunately, any Ising quantum annealer today is incapable of embedding any out-of-plane terms in the Hamiltonian which are required to enumerate the out-of-plane orders observed in $HoB_4$. It turns out that a complete treatment of the full Hamiltonian may require some of these computing methods to come of maturity in the future. Yet, the data offers a compelling case for a machine-learning based Hamiltonian discovery, where AI kernels could be trained using not only classical, [43] but also quantum treatments of the Hamiltonians, both with and without defects, and then tested on our dataset here.

### VII. Acknowledgement

We thank Akshat A. Jha for his initial computational contribution. This research is funded by the U.S Department of Energy (US-DOE) – Office of Science, Basic Energy Sciences Grant No. DE-SC0022986, under the project "Seeking Quasiparticles from Low-Energy Spin Dynamics." A portion of the research (funding of G.K.) comes from US-DOE national Quantum Initiative Science Research Center - Quantum Science Center. H.X. and the work with D-Wave, are supported by the Center for Quantum Technologies (CQT) under the industry-University Cooperative Research Center (IUCRC) Program at the US National Science Foundation (NSF) under Grant No. 2224960. The authors acknowledge support from Neil Dilley, and the support of spin lab in BIRCK facility of Purdue University. A portion of this research used resources at the Spallation Neutron Source, a DOE Office of Science User Facility operated by the Oak Ridge National Laboratory. The beam time was allocated to BL-9 CORELLI on proposal number IPTS-24250.1.